\def\vecr{\vec r}
\def\vecp{\vec p}
\def\Ntest{N_{\textrm{test}}}
\def\Maria{}
\documentclass{camera}
\usepackage{graphicx}\usepackage{epsfig}\usepackage{portland}

\begin{document}

%
\title{Boltzmann-Langevin One-Body\\
dynamics for fermionic systems}

%
\author{P. Napolitani$^1$ \and M. Colonna$^2$}

%
\organization{\vspace{-1.5cm}
$^1$ IPN, CNRS/IN2P3, Universit\'e Paris-Sud 11, 91406 Orsay cedex, France\\
$^2$ INFN-LNS, Laboratori Nazionali del Sud, 95123 Catania, Italy}

\maketitle
\vspace{-0.5cm}
\begin{abstract}
	A full implementation of the Boltzmann-Langevin equation for
fermionic systems is introduced in a transport model for dissipative 
collisions among heavy nuclei.
	Fluctuations are injected in phase space and not,
like in more conventional approaches, as a projection on
suitable subspaces.
	The advantage of this model is to be specifically adapted to
describe processes characterised by instabilities, like the formation
of fragments from a hot nuclear system, 
and by dissipation, like the transparency in nucleus-nucleus collisions.
\end{abstract}
 
\section{Fluctuations and bifurcations}
%
%
	Dissipative nucleus-nucleus collisions are a unique probe
for the in-medium nuclear interaction at densities away from 
saturation and at high nucleon momenta.
	In weakly-excited systems, Pauli-blocking factors of final orbitals largely 
suppress two-body direct interactions.
	In a semiclassical framework, the dynamics of such situation can 
be appropriately described within a Vlasov formalism, where
the temporal evolution of the one-body distribution function $f(\vecr,\vecp,t)$ 
(function of time $t$, space coordinates $\vecr$ and momentum $\vecp$)
in the self-consistent one body field is governed by the effective Hamiltonian 
$H[f]$.
	However, for more violent collisions, direct two-body
interactions become significant and require to be treated by an additional
Boltzmann collision integral $I[f]$.
	In the Uehling-Uhlenbeck form, this latter is introduced 
as a continuous-source term $\bar{I}[f]$.
	Within standard forms of BUU/VUU/BNV formalisms, the additional
use of an ensemble-averaged mean field makes the transport model
particularly suited for one-body observables 
(for example the collective flow, or proton spectra),
but suppresses bifurcations.

	Bifurcations are however crucial in processes characterised by volume
instabilities, where fluctuations are the seeds of the formation of
several intermediate-mass fragments. 
	A possible approach to this scenario is the employment of a two-body 
Hamiltonian, as in Molecular Dynamics~\cite{AichelinOno}.
	As an extension of the one-body dynamics, another approach is the introduction
of one more term $\delta I[f]$ which produces Langevin fluctuations.
The resulting temporal evolution of the one-body distribution function 
in a Boltzmann-Langevin approach is given by
\begin{equation}
	\partial_t\,f= \left\{H[f],f\right\}+{\bar{I}[f]}+{\delta I[f]} \;.
\label{eq1}
\end{equation}
	This latter is the method we follow.


	Many approaches to the Boltzmann-Langevin equation in nuclear transport models are 
projections on one suitable subspace, like the {\Maria coordinate} space.
	In this framework, a fluctuating term can be prepared by associating a Brownian force to a stochastic one-body
potential~\cite{Guarnera1997} or it can correspond to kinetic equilibrium fluctuations of 
a Fermi gas~\cite{Colonna1998}.
	The disadvantage of projection methods is that the amplitude of the fluctuation may have 
to be adjusted and fluctuations may appear from a specific time on (when equilibrium is attained 
or when entering spinodal conditions), as an amplification of the most unstable mode.

	The approach we follow consists in constructing a stochastic collision term which includes 
Boltzmann-Langevin fluctuations in phase space, supplemented by a strict treatment 
of Pauli-blocking factors, in the framework of semiclassical test-particle-based transport models.

\section{Collision statistics and fragment formation}
%
%
\begin{figure}[b!]\begin{center}
	\includegraphics[angle=0, width=0.87\textwidth]{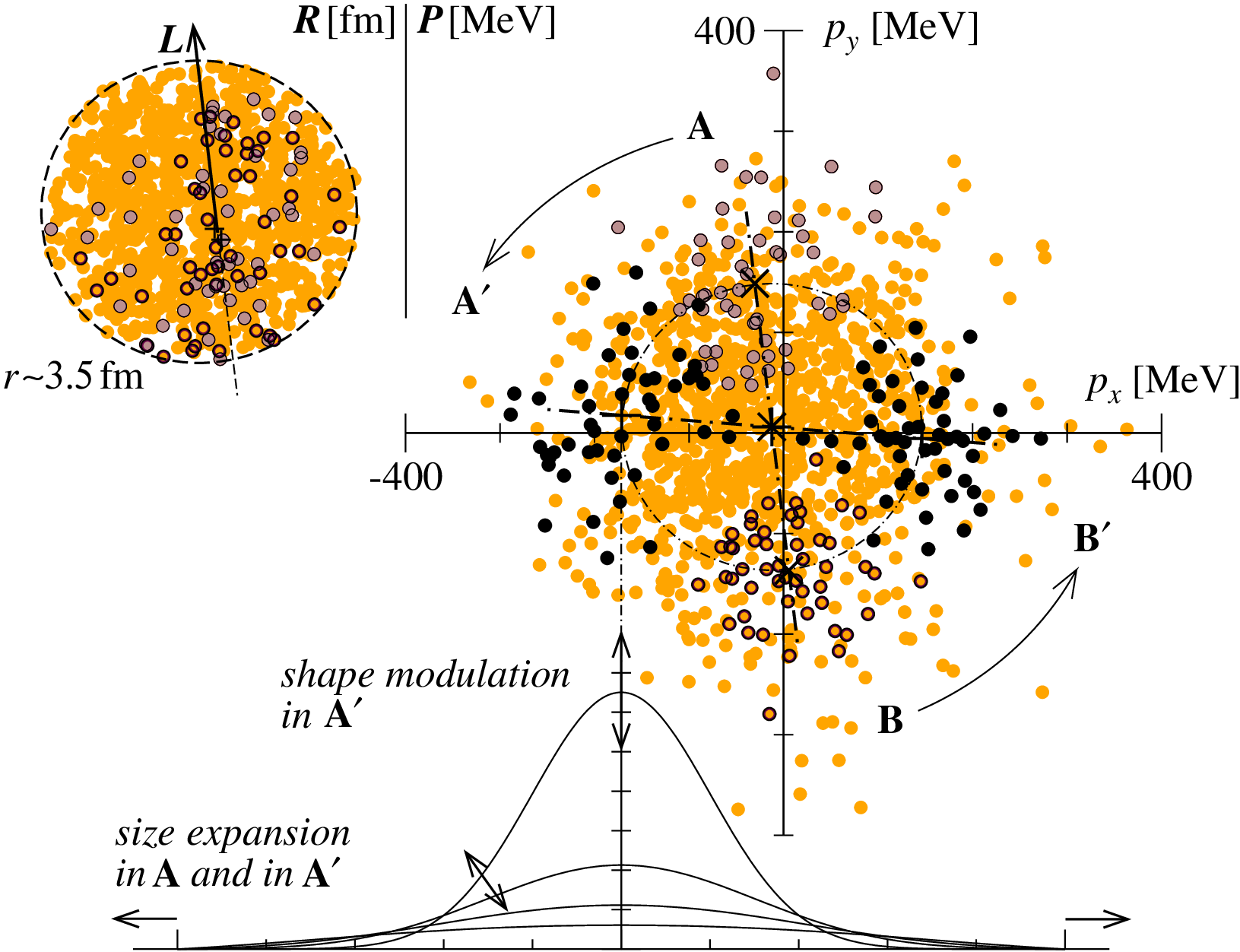}
\end{center}\caption
{
	Example of one collision event in BLOB.
	Two nucleons are represented by two agglomerates of test-particles $A$ and $B$ which
share the same volume in coordinate space $\mathcal{R}$. In the corresponding momentum
space $\mathcal{P}$ the collision of $A$ and $B$ originates by rotation (determined by the
scattering angle) the destination sites $A'$ and $B'$, where the test-particles are 
redistributed according to Pauli-blocking and energy constraints.
%
%
}
\label{fig_vortex}
\end{figure}
%
%
\begin{figure}[b!]\begin{center}
\includegraphics[angle=0, width=1\textwidth]{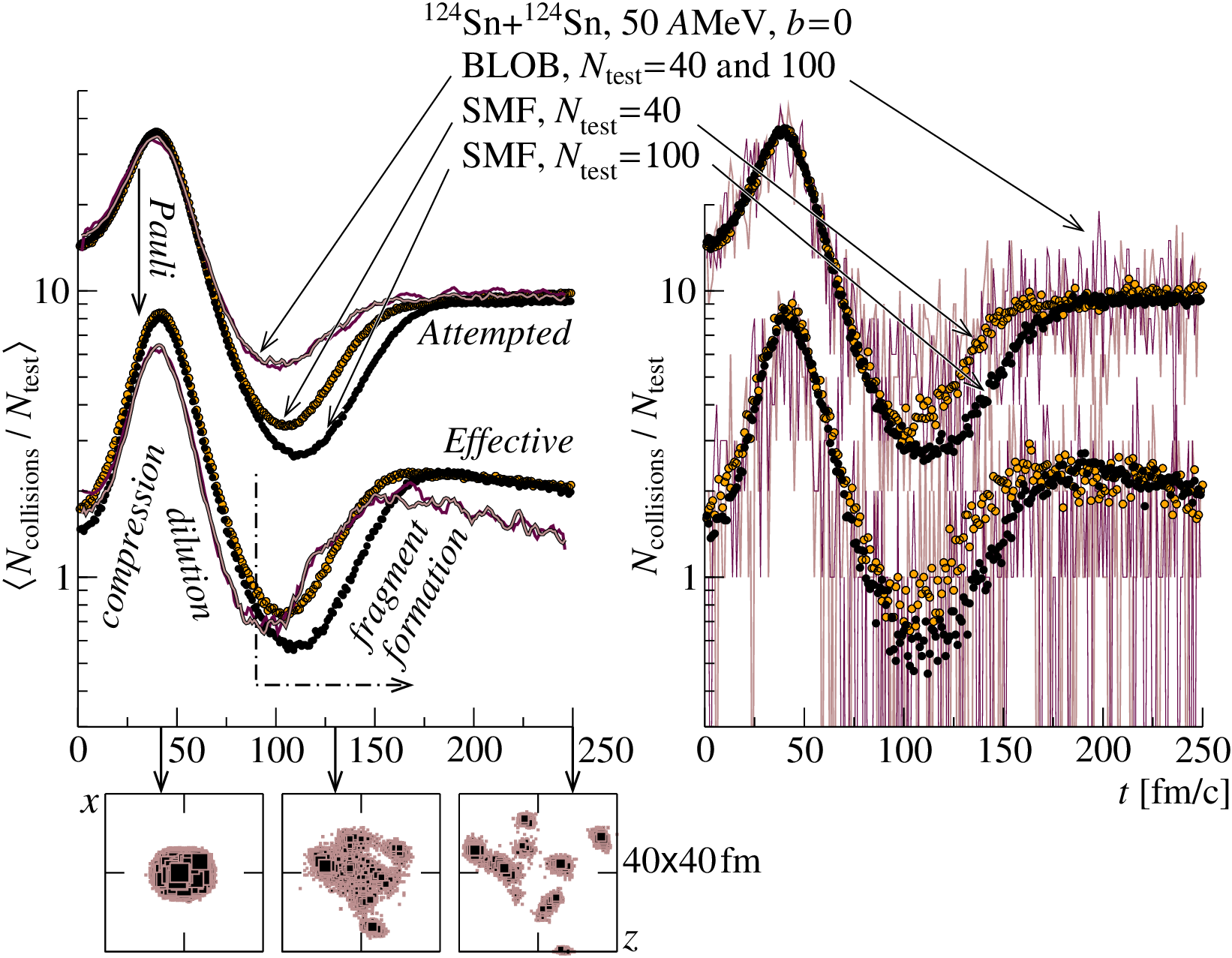}
\end{center}\caption
{
	SMF and BLOB compared for the same system $^{124}$Sn$+^{124}$Sn at 50 $A$MeV and $b=0$.
	Left. Mean value of the evolution of the number of collisions 
extracted from a bundle of trajectories.
	Right. Amplitude of the fluctuations shown by the evolution of the number 
of collisions for single events.
}
\label{fig_fluctuations}
\end{figure}
%
%

	We construct the Bolzmann-Langevin one body (BLOB) model by using the same 
semiclassical isospin-dependent mean-field employed in the Stochastic Mean Field (SMF) 
model~\cite{Guarnera1997,Chomaz2004}; this is characterised by an incompressibility of
$k_{\inf}=200$ MeV (soft) and the potential component of the symmetry energy is given 
either by a linear term as a function of the density (asy-stiff),
or by a quadratic term, {\Maria with negative curvature} (asy-soft).

%
%
%
%
{\Maria In SMF, the collision term is introduced in the standard Uehling-Uhlenbeck form and fluctuations are injected 
in a reduced subspace, agitating the spatial density profile.  On the other hand, in BLOB 
fluctuations are implemented in full phase-space, replacing the collision integral by the Boltzmann-Langevin 
approach described in the following.}
	Like in the method {\Maria introduced} in ref.~\cite{Bauer1987}, the collision term of BLOB 
involves entire nucleons.
	The test-particle method is used and a number of $\Ntest$ test-particles per nucleon 
is employed.
	A `nucleon' is represented by an agglomerate of $\Ntest$ test-particles of identical 
isospin which share the same volume in coordinate {\Maria and in momentum} space, as shown in fig.~\ref{fig_vortex}.

	The number of attempted collisions is weighted on the probability of satisfying the 
mean-free-path condition for the two agglomerates of test particles 
in a given interval of time.
	Such probability is proportional to the mean density in momentum space,
the average relative momentum between the two clouds and the nucleon-nucleon cross section.
	This latter is divided by a factor $\Ntest$ in order to keep a correspondence with the 
Uehling-Uhlenbeck collision term.
	In fig.~\ref{fig_fluctuations}, SMF and BLOB are used to simulate the 
same system $^{124}$Sn$+^{124}$Sn at 50 $A$MeV and $b=0$.
	We use here a simplified version of the SMF model, where fluctuations in the phase-space mapping are generated
by the use of a finite number of test particles, $N_{test}$, in the numerical resolution of Eq.~\ref{eq1}.
Then, the amplitude of fluctuations injected in the dynamics scales with $1/N_{test}$.
	Several events were simulated so that several trajectories could be drown for the evolution of the number of 
attempted and effective collisions.	
	Mean trajectory could be deduced as the average of the bundle
of all the simulated trajectories, this is shown on the left panel; the right panel shows one single trajectory 
per type of calculation so as to appreciate the amplitude of the fluctuation.
	The evolution of the number of collisions as a function of time is a probe of the reaction mechanism.
	This number increases during the initial stage of the collision, which is a phase of compression.
	A stage of dilution follows, characterised by a drop of the number of collisions.
	The trend then reaches a minimum and inverses signing the beginning of the process of fragment formation, 
characterised by an initial increase of the number of collisions and a successive stabilisation and levelling off
due to the separation of the fragments.
	This figure shows that, as expected, the number of attempted collisions per test particle 
averaged over a statistics of dynamical trajectories should not differ in the present 
formalism with respect to a conventional BUU approach when the initial rising side of the 
spectrum is considered.
	On the other hand, the amplitude of the fluctuation marks the difference between the two models.
	In correspondence with the phase of dilution (around 80-120 fm/c), fluctuations characterise 
not only the statistics of collisions but also the kinematics.
	The simulation was repeated for $N_t=40$ and $N_t=100$.
	The incomplete treatment of fluctuations used in SMF results in establishing a dependence of the collision statistics
on the numerical parameter $N_t$.
	When $N_t$ is reduced in SMF the fragment formation is anticipated, the dynamics becomes more explosive, 
and the fragment multiplicity increases.
	All these effects are even further enhanced when moving from SMF with $N_t=40$ to BLOB, but in the latter
case the dependence on $N_t$ disappears,  
{\Maria as expected, because in BLOB a fluctuation-source term, corresponding to the Boltzmann-Langevin collision integral, has been
implemented.}  

%
%
\begin{figure}[b!]\begin{center}
\includegraphics[angle=0, width=.72\textwidth]{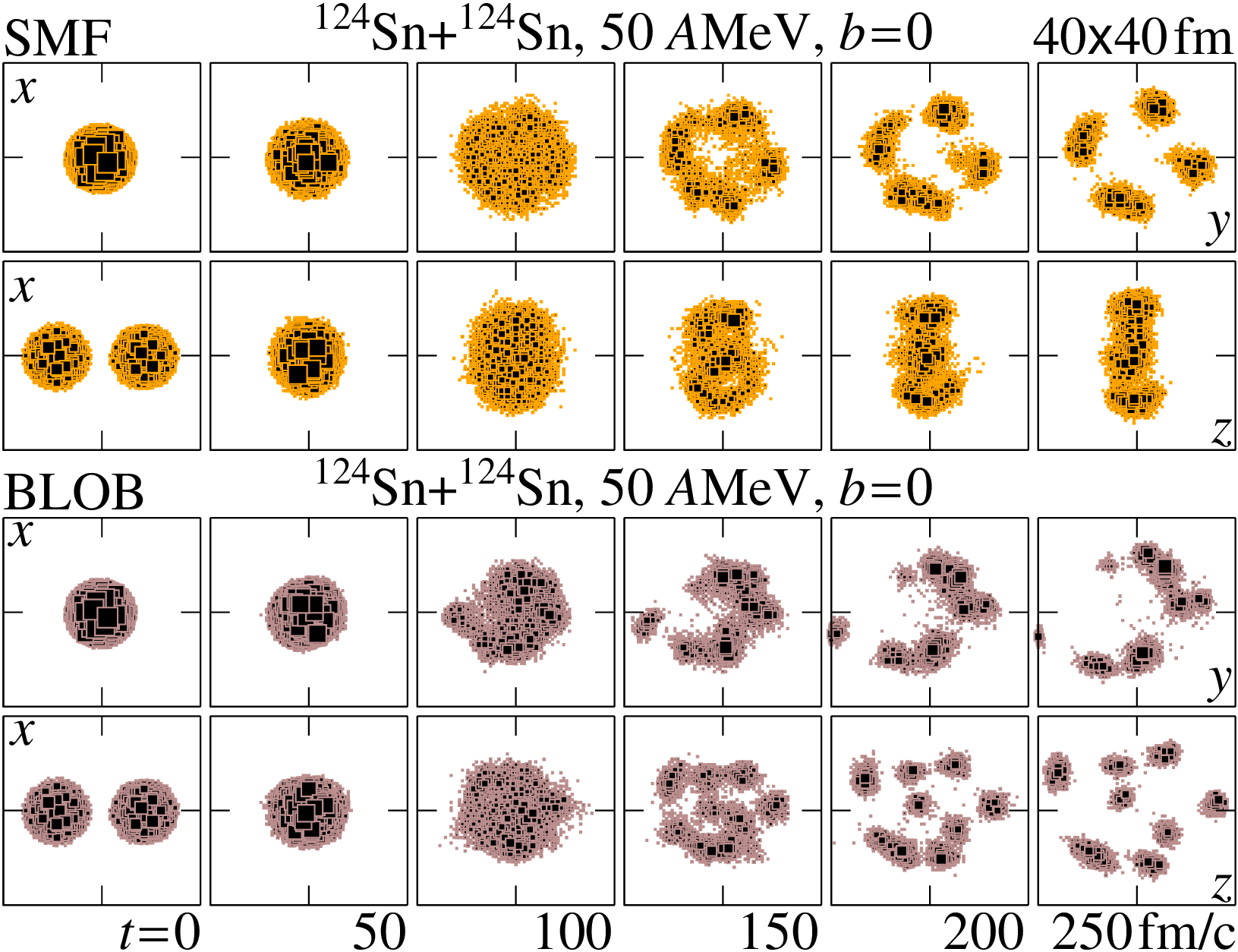}\\\vspace{1ex}
\includegraphics[angle=0, width=.72\textwidth]{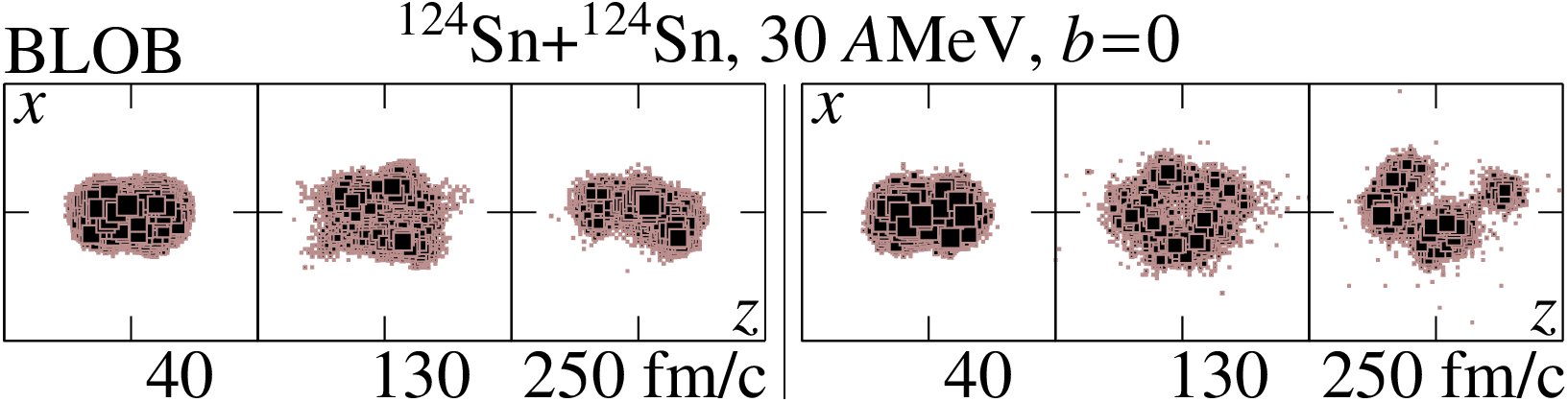}
\end{center}\caption
{
	First and second rows. SMF and BLOB compared for the same system $^{124}$Sn$+^{124}$Sn at 50 $A$MeV and $b=0$.
	$N_t=100$ is used.
	Third row. Two different exit channels for  the same system 
	$^{124}$Sn$+^{124}$Sn at 30 $A$MeV and
	$b=0$, calculated with BLOB.
}
\label{fig_film}
\end{figure}
	Like in the approach followed in ref.~\cite{Rizzo2008}, the effective-collision 
probability is weighted on the blocking factors of the destination sites, 
taking into account the full extension of the initial distributions.
	In BLOB, distributions are adapted to the occupation profile
by the use of a numerical optimisation procedure.
	This latter converges first to the appropriate extension of the scattered regions
(by expanding initial and final sites in momentum space) and then converges
to the appropriate test-particle final distribution by reorganising the scattered test
particles according to packets which can be modulated in shape; the principle of this
method is shown in fig.~\ref{fig_vortex}.
	Due to the inclusion of blocking factors in the collision probability, the number of 
effective collisions drops to smaller values with respect to the number of attempted 
collisions.
	The evolution of the effective-collision number is not expected to be identical for 
SMF and BLOB calculations; the number is reduced with respect to the 
attempted-collision spectrum, keeping the shape mostly unchanged, but the effective 
collisions result more penalised by the Pauli blocking treatment in BLOB then in SMF.
	On the one hand, this difference comes from the fact that the Pauli blocking 
is applied to the scattering of test-particle couples in SMF, while in BLOB it is applied 
to extended distributions of test particles. 
	On the other hand, it is important to remark that in BLOB the number of effective 
collisions is reduced with respect to SMF but in this latter case collisions are more 
efficient in producing phase-space fluctuations.
	Moreover, in the BLOB calculation the fluctuation is not gradually amplified 
(the SMF approach results in amplifying the most unstable mode) but it presents large and 
correct amplitudes since the initial instants: this favours the early appearing of 
bifurcations and explains the earlier fragment formation and the larger kinetic energy 
of the fragments.
	As a further outcome, a larger variety of exit-channel configurations are expected.

%
%
	In fig.~\ref{fig_film}, like in fig.~\ref{fig_fluctuations},
SMF and BLOB are again used to simulate the system $^{124}$Sn$+^{124}$Sn at 50 $A$MeV and $b=0$.
	In full agreement with the analysis of the collision statistics and the properties of fragments,
we find larger fragment kinetic energies for the BLOB simulation.
	The fragment configuration is more isotropic in the BLOB simulation, this could be
due to a larger transparency.
A similar comparison was done between SMF and AMD in ref.~\cite{RizzoColonna}.

Another calculation is shown in fig.~\ref{fig_film} for the less excited system $^{124}$Sn$+^{124}$Sn at 30 $A$MeV, 
close to the threshold between fusion and fragmentation.
This analysis has the purpose of studying the effect of bifurcation.
In the present example we find that in the SMF calculation one trajectory prevails (leading to fusion), 
while in the BLOB calculation a more complete phase-space sampling allows to access 
a variety of exit-channel configurations (from fusion to fragmentation).

\section{Conclusions}	

	The differences between BLOB and SMF appear in the statistics of collisions as well as 
in the kinematics.

	In this context, the study of a new strategy to solve the 
Boltzmann-Langevin equation is presented.
	As an extension of the Bauer-Bertsch approach, in the framework 
of semiclassical test-particle-based transport models, fluctuations 
of correct amplitude are introduced in phase space through a stochastic 
collision term. 
	The resulting fluctuations have so large amplitude to induce 
bifurcations in the dynamical paths of the one-body phase-space density 
and they correlate over smaller volumes with respect to previous forms 
of the collision integral.
	In prospective, this strategy is promising for the improvement of 
transport models for nuclear collisions and for the implementation of 
isovector effects in the collision dynamics.

%
%
%
%


\begin{thebibliography}{99}
%
\bibitem{AichelinOno}
J. Aichelin, Phys. Rep. 202 (1991) 233; \\
A. Ono, Phys. Rev. C 59 (1999) 854.
%
\bibitem{Guarnera1997}
A. Guarnera et al.
Phys. Lett. B403 (1997) 191.
%
\bibitem{Colonna1998}
M. Colonna et al.
Nucl. Phys. A642 (1998) 449.
%
\bibitem{Chomaz2004}
Ph. Chomaz, M. Colonna and J. Randrup,
Phys. Rep. 389 (2004) 263.
%
\bibitem{Bauer1987}
W. Bauer and G.F. Bertsch,
Phys. Rev. Lett. 58 (1987) 863
%
\bibitem{Rizzo2008}
J. Rizzo, Ph. Chomaz and M. Colonna,
Nucl. Phys. A (2008).
%
\bibitem{RizzoColonna}
J. Rizzo, M. Colonna, A. Ono, Phys. Rev. C 76 (2007) 024611;\\
M. Colonna, A. Ono, J. Rizzo, Phys. Rev. C 82 (2010) 054613.
%
\end{thebibliography}
\end{document}